\documentclass[conference,a4paper]{IEEEtran}
\IEEEoverridecommandlockouts

\usepackage[dvips]{graphicx}
\usepackage{fancyhdr}
\usepackage{eclbkbox}
\usepackage{indent}
\usepackage{subfigure}
\usepackage{url}
\makeatletter
\def\tbcaption{\def\@captype{table}\caption}
\def\figcaption{\def\@captype{figure}\caption}

\usepackage{fancyhdr}

\begin{document}
\title{A Recommendation System of Grants\\ to Acquire External Funds
\thanks{\copyright 2016 IEEE. Personal use of this material is permitted. Permission from IEEE must be obtained for all other uses, in any current or future media, including reprinting/republishing this material for advertising or promotional purposes, creating new collective works, for resale or redistribution to servers or lists, or reuse of any copyrighted component of this work in other works.}
}

\author{
\IEEEauthorblockN{Shin Kamada}
\IEEEauthorblockA{Dept. of Intelligent Systems\\
Graduate School of Information Sciences\\
Hiroshima City University\\
3-4-1, Ozuka-Higashi, Asa-Minami-ku\\
Hiroshima, 731-3194, Japan\\
da65002@e.hiroshima-cu.ac.jp}
\and
\IEEEauthorblockN{Takumi Ichimura}
\IEEEauthorblockA{Community Liaison Center \&\\
\ Faculty of Management and Information Systems\\
Prefectural University of Hiroshima\\
1-1-71, Ujina-Higashi, Minami-ku\\
Hiroshima, 734-8559, Japan\\
ichimura@pu-hiroshima.ac.jp}
\and
\IEEEauthorblockN{Takanobu Watanabe}
\IEEEauthorblockA{Community Liaison Center\\
\\
Prefectural University of Hiroshima\\
1-1-71, Ujina-Higashi, Minami-ku\\
Hiroshima, 734-8559, Japan\\
t-watanabe@pu-hiroshima.ac.jp}
}

\maketitle

\pagestyle{fancy}{
\fancyhf{}
\fancyfoot[R]{}}
\renewcommand{\headrulewidth}{0pt}
\renewcommand{\footrulewidth}{0pt}

\begin{abstract}
The recommendation system of the competitive grants to university researchers by using the Grants-in-Aid for Scientific Research (KAKEN) keywords has been developed. The system can determine the recommendation order of researchers to each grant by the using the association rules between KAKEN application and various information from the web site of the corresponding grant. However, our developed previous system has some fatal errors in the retrieval algorithm. We modify the algorithm and extend the retrieval data for web mining. If the grant information is not enough to determine the relation, the system investigates the past KAKEN records in the database for the researcher who acquired the past grant. Moreover, the system retrieves the papers of the researchers to search their interests. As a result, the agreement degree of the researcher's interest to the grant increases. This paper discusses some simulation results.
\end{abstract}

\begin{IEEEkeywords}
Recommendation System, Grants-in-Aid, Association Analysis, TF-IDF, Web Intelligence
\end{IEEEkeywords}

\IEEEpeerreviewmaketitle

\section{Introduction}
With regard to expenditures for education and science, Japan government will promote reforms aiming to improve the quality of education and research as well as improve the quality of relevant budgets by increasing competitive research grants. The phenomenon about decrease of research expense appeared in the various university. For the avoidance of that problem, all researchers in the university are going to endeavor to acquire external research funds. Grants-in-Aid for Scientific Research (KAKEN) Program \cite{KAKENHI} is one of the most popular funds provided by Japan governments, called Ministry of Education, Culture, Sports, Science and Technology (MEXT) and Japan Society for the Promotion of Science (JSPS). Therefore, the university researcher proposes an application of his/her research plan of KAKEN based on the classification table of KAKEN keywords. KAKEN keywords provide various kinds of fundamental one for each research field and they are updated once a year according to the current research trends. However, the achievement of acquisition of KAKEN is not easy, therefore the researcher may have to challenge the additional grants provided by non-governmental organizations. 

For such a situation, we have developed the recommendation system with the matching degree between the university researchers and the grant provided by non-governmental organizations by using KAKEN keywords \cite{Kamada15}. The system can determine the recommendation order of researchers to each grant by using the association rules \cite{Vapnik95} between KAKEN application and various information from the web site of the corresponding grant. The reason to use KAKEN keywords is that each grant has the different information or original keywords at web site, we accepted the idea to use KAKEN keywords to discovery another competitive grants as the main key. However, the system had some fatal errors in the retrieval algorithm. The system cannot determine the recommendation order successfully when the grant information is sufficiently too small to discover them. We have modified the algorithm and extend the retrieval data for web mining. If the grant information is not enough to determine the relation, and the system investigates the past records in KAKEN database. Moreover, the system retrieves the past written papers of the researchers to implement more informative searches. As a result, the agreement degree of the researcher's interest to the grant increases. We report some simulation results by using the developed system in this paper.

The remainder of this paper is organized as follows. Section \ref{sec:DevelopedSystem} describes our revised recommendation system. Section \ref{sec:SyntheisEvaluation} describes the evaluation process in case of the use of past paper information. Section \ref{sec:ExperimentalResults} describes experimental results for some competitive grants. In Section \ref{sec:Conclusion}, we give some discussions to conclude this paper.

\section{Recommendation System of Grants}
\label{sec:DevelopedSystem}
\subsection{System Overview}
\label{sec:system}

\begin{figure*}[tbp]
\begin{center}
\includegraphics[scale=0.9]{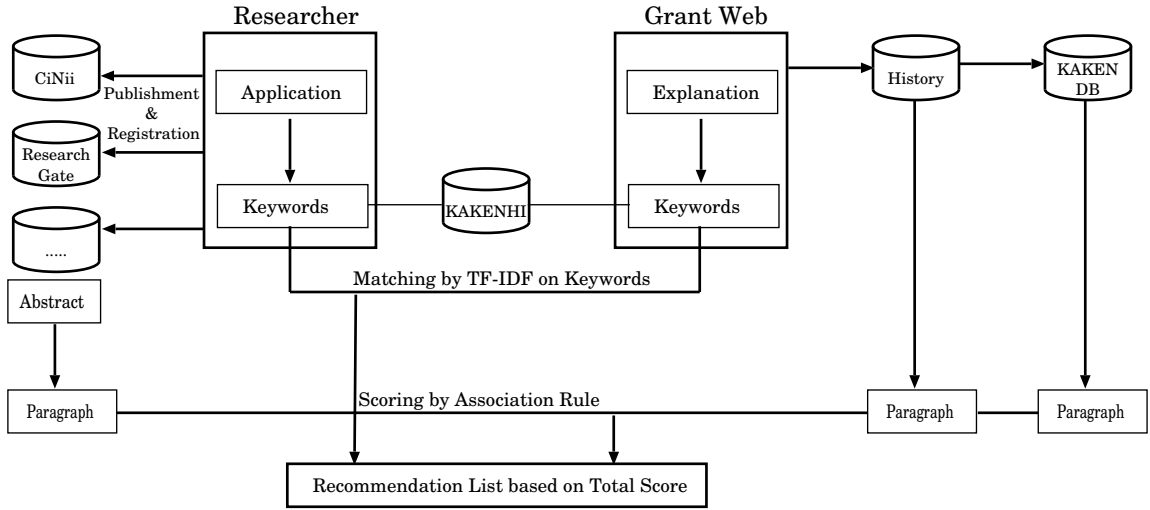}
\vspace{-3mm}
\caption{The overview of the recommendation system}
\label{fig:overview}
\end{center}
\end{figure*}

In \cite{Kamada15}, we developed the recommendation system of Grants-in-Aid for researchers which the agreement degree between the current research theme of the researchers and the Grant information is calculated by using the KAKEN keywords table \cite{KAKENHI}. The system in \cite{Kamada15} made the association rules between the information of Grant site and KAKEN keywords, and the matching system of the researchers and the grant by the keyword was constructed. However, the matching for the highest order of researchers was performed. The improvement of system should be required to build the association between the researchers and the Grants.

Fig. \ref{fig:overview} shows the overview of the revised system. The system consists of 2 analysis parts. The simple analysis algorithm checks the surface data such as the application of Grant web site. The part of system executes the matching degree between the keywords in researcher's application of KAKEN and the information on Grant Website by using the KAKEN keywords. The latter part checks the agreement degree by using the association rules which analyzed the researcher written papers from the web database such as the CiNii database \cite{CiNii}, the Research gate database \cite{ResearchGate}, and the historical data such as selection results. The historical data is mined from the accepted KAKEN's abstract of the selected researcher with the retrieval from KAKEN database.

\subsection{KAKEN Keywords \cite{JSPS-keyword}}
\label{sec:JSPSKeyword}
This section explains the KAKEN keywords briefly. The KAKEN keywords are classified the researcher fields into large groups, medium groups, and small groups to construct a hierarchical retrieval structure.

\begin{table*}[tbp]
\caption{An example of KAKEN Keywords \cite{JSPS-keyword}}
\label{tab:JSPSkeyword}
\vspace{-3mm}
\begin{center}
\scalebox{1.0}[1.0]{
\begin{tabular}{c|c|c|l}
\hline \hline
Category       &  Sub Category              &  Field              & \multicolumn{1}{c}{Keyword}\\ \hline\hline
Informatics  &  Principles of Informatics   &  Theory of informatics       & Theory of computation, Automata theory, $\cdots$\\
             &                              &  Mathematical informatics    & Optimization theory, Mathematical finance, $\cdots$\\
             &  Human informatics           &  Intelligent informatics     & Machine learning, Knowledge acquisition, $\cdots$\\ \hline
Complex systems &  Human life science       &  Eating habits               & Cooking and processing, Food storage, $\cdots$\\
&                           &  Clothing life/Dwelling life & Dwelling life, Clothing culture, $\cdots$\\ \hline
Engineering & Mechanical engineering      &  Materials/Mechanics of materials               & Continuum mechanics, Structural mechanics, $\cdots$\\
&                           &  Thermal engineering & Thermophysical property, Convection, $\cdots$\\ 
&   Integrated engineering & Aerospace engineering& Aerodynamics, Structure/Material, $\cdots$\\ \hline
$\cdots$ &  $\cdots$  &  $\cdots$ & $\cdots$\\
\hline \hline
\end{tabular}
} 
\end{center}
\end{table*}

Table \ref{tab:JSPSkeyword} shows a part of KAKEN keyword table, which consists of 4 columns: `Category', `Sub Category', `Field', and `Keyword'. The term of `Category,' `Sub Category,' and `Field' indicate the hierarchical research category. Keyword is the words of explanation, technical term, and so on. The table has 14 Categories, 80 Sub Categories, 322 Fields, and 3674 Keywords, respectively. The Research Filed has about 11 keywords on average.

\subsection{Web Data Mining}
\label{sec:Webdatamining}
The subsection explains web data mining that is the techniques of the extraction of digital text files from web site of each grant organization. Each grant organization provides the original content in details and the historical results related to the grant, because they promote the increase of the number of applications. For each grant organization, we extract the text files as follows.
\begin{enumerate}
\item Download grant information\\
\label{enum:Webdata_1}
First of all, PDF and HTML files are downloaded from Web site in each grant organization by using ``wget''.

\item Convert to text files\\
HTML files and PDF files that obtained in Step \ref{enum:Webdata_1}) are converted to text files. The following regular expression is used to remove HTML tags and the redundant information.
\begin{verbatim}
<("[^"]*"|'[^']*'|[^'">])*>
\end{verbatim}

In case of the PDF file, the process of the text extraction is executed by using ``pdftotext'' or ``poppler''.
\end{enumerate}

\subsection{TF-IDF}
\label{sec:TF-IDF}
The TF-IDF calculates a weight often used in information retrieval and text mining. The term frequency $tf(t,d)$ gives a measure of the importance of the term $t$ within the particular document $d$. The inverse document frequency $idf(t)$ is a measure of the general importance of the term. A high weight in $tfidf$ is reached by a high term frequency and a low document frequency of the term in the whole collection of documents.

\begin{eqnarray}
\nonumber tf(t, d)&=&\frac{n(t,d)}{\sum_{k} n(k,d)}\\
\nonumber idf(t) &=& \log \frac{|D|}{|\{d \in D: t \in d\}|}\\
tfidf(t,d)&=&tf(t,d) \times idf(t),
\label{eq:tfidf}
\end{eqnarray}
where $n(t,d)$ is the occurrence count of a term $t$ in the document $d$. $|D|$ is the total number of documents in the corpus. $|\{d \in D: t \in d\}|$ is the number of documents where the term $t$ appears.

In this paper, $d$ and $t$ indicate the html file in Grant web site and the representative paragraph in the explanation, respectively.

\subsection{Association Rule}
\label{sec:SystemAssociationRule}
Many machine learning algorithms for the data mining are developed. Support Vector Machine \cite{Vapnik95} is one of mathematical algorithms to data science work with numeric data. Association rule is the perfect data mining method for non numeric data. Association rule is focused on the discovery of frequent co-occurrence among many items. Association rule mining is try to find frequent association rules between items with strong relationships. We give a brief reminiscent explanation in this section.

Each rule as shown in Table \ref{tab:transaction} is represented as follows \cite{TAN05}. Table \ref{tab:transaction} illustrates commonly known as market basket transactions.

\begin{equation}
\{X\} \rightarrow \{Y\},
\label{eq:associationrule1}
\end{equation}
where $X$ and $Y$ are items such as `Machine Learning', `Neural Network'. The rule as shown in Eq.(\ref{eq:associationrule1}) has the existence of strong relationship between $X$ and $Y$. That is, if $X$ occurs then $Y$ also occurs. 

\begin{table}[tbp]
\caption{An example of database}
\vspace{-3mm}
\label{tab:transaction}
\begin{center}
%\scalebox{1.2}[1.2]{
\begin{tabular}{c|p{5cm}}
\hline
Transaction ID  & Item Sets\\ \hline
1  &  \{Machine Learning, Neural Network\} \\\hline
2  &  \{Machine Learning, Information Retrieval, Knowledge Acquisition, Industrial Engineering\} \\ \hline
3  &  \{Neural Network, Information Retrieval, Knowledge Acquisition, Information Theory\} \\ \hline
4  &  \{Machine Learning, Neural Network, Information Retrieval, Knowledge Acquisitions\} \\ \hline
5  &  \{Machine Learning, Neural Network Information Retrieval, Information Theory\} \\ \hline
\end{tabular}
%} 
\end{center}
\end{table}

Let $\boldmath{I}=\{i_{1}, i_{2}, \cdots, i_{d} \}$ be the set of all items and $\boldmath{T}=\{t_{1}, t_{2}, \cdots, t_{N} \}$ be the set of all transactions. Each transaction $t_{i}$ contains a subset of items from $\boldmath{I}$. In association analysis, a collection of zero or more items is called an item set. The width of transaction is the number of items in a transaction. The transaction $t_{j}$ contains an item set $X$ if $X$ is a subset of $t_{j}$. The support count is the important property of item set. The support count refers to the number of transactions including a particular item set. The support count for an item set $X$ can be expressed by Eq.(\ref{eq:associationrule2}).
\begin{equation}
  \sigma(X)=|\{t_{j}|X \in t_{j}, t_{j} \in T\}|
\label{eq:associationrule2}
\end{equation}

The association rule is an implication of the $X \rightarrow Y$, where $X \cap Y = \phi$. It is natural that the strength of association rule is given as the support and the confidence. The support is how often the rule is applicable to a data set and the confidence is how frequency items in the consequent part $Y$ appear in the transaction including $X$ as follows.
\begin{equation}
{\rm supp}(X \rightarrow Y)=\frac{\sigma (X \cup Y)}{N},
\label{eq:support}
\end{equation}
\begin{equation}
  {\rm conf}(X \rightarrow Y)=\frac{\sigma(X \rightarrow Y)}{\sigma(X)}\\
  =\frac{{\rm supp}(X \rightarrow Y)}{{\rm supp}(X)},
\label{eq:confidence}
\end{equation}

Moreover, the lift is one more parameter of interest in the association analysis. The lift is nothing but the ratio of confidence to expected confidence as follows.
\begin{equation}
{\rm lift}(X \rightarrow Y)=\frac{{\rm conf}(X \rightarrow Y)}{{\rm supp}(Y)},
\label{eq:lift}
\end{equation}

If the support is low, the probability that the rule appears is low. However, in case of big data analysis where there are many kinds of items, we cannot expect that the support of each item is large. Therefore, the evaluation of association rule needs the support, confidence, and lift.

In our developed system, the association rules are retrieved as follows.
\begin{enumerate}
\item The extraction of sentence in an abstract of science paper by the regular expression. The sentence is a transaction of association rule.
\label{enum_SystemAssociationRule_1}
\item The Japanese morphological analysis such as MeCab \cite{Mecab} is analyzed for a transaction and then nouns are extracted. The noun is an item and make the item set for the transaction.
\label{enum_SystemAssociationRule_2}
\item The historical data the selected results from the Grant web site are extracted and then the procedure from Step \ref{enum_SystemAssociationRule_1}) to Step \ref{enum_SystemAssociationRule_2}) is executed. As a result, the set of transaction rules and item sets are calculated.
\label{enum_SystemAssociationRule_3}
\item Marge the transaction rules and the item set in Step \ref{enum_SystemAssociationRule_2}) and Step \ref{enum_SystemAssociationRule_3}) and the association rules are calculated by the method described above.
\label{enum_SystemAssociationRule_4}
\item Select the rules related to the researchers in the university.
\label{enum_SystemAssociationRule_5}
\item Calculate the score by the association rules in Step \ref{enum_SystemAssociationRule_3}).
\label{enum_SystemAssociationRule_6}
\end{enumerate}

\begin{table*}[tbp]
\caption{Recommendation Lists and Matching Score of Surface data by KAKEN keywords on \cite{kayamori}}
\vspace{-3mm}
\label{tab:result-keyword-kayamori}
\begin{center}
\scalebox{1}[1]{
\begin{tabular}{l|r|r|p{7cm}}
\hline \hline
\multicolumn{1}{c|}{Researcher Name} & \multicolumn{1}{c|}{No. of Matched KAKEN keywords} & \multicolumn{1}{c|}{Surface data} & \multicolumn{1}{c}{Matched KAKEN keywords} \\ \hline\hline
Researcher 1-A  & 3 & 0.708 & `Information Retrieval',  `Natural Language Processing',  `Knowledge Acquisition'  \\
Researcher 1-B  & 2 & 0.608 & `Information Theory,  `Industrial Engineering' \\
Researcher 1-C  & 2 & 0.377 & `Machine Learning,  `Neural Network' \\
Researcher 1-D  & 2 & 0.350 & `Knowledge Acquisition,  `Neural Network' \\
Researcher 1-E  & 2 & 0.250 & `Neuroinformatics,  `Computational Neuroscience' \\
\hline \hline
\end{tabular}
} 
\end{center}
\end{table*}

\begin{table*}[tbp]
\caption{Recommendation Lists and Matching Score of Historical data by Association Rule on \cite{kayamori}}
\vspace{-3mm}
\label{tab:result-rule-kayamori}
\begin{center}
\scalebox{1}[1]{
\begin{tabular}{l|r|r|p{7cm}}
\hline \hline
\multicolumn{1}{c|}{Researcher Name} & \multicolumn{1}{c|}{No. of Matched Association Rules} & \multicolumn{1}{c|}{Historical data} & \multicolumn{1}{c}{Matched Association Rules} \\ \hline\hline
Researcher 1-C  & 2 & 0.759 & \{Reinforcement Learning\} $\rightarrow$ \{Machine Learning\}, \{Reinforcement Learning\} $\rightarrow$ \{Neural Network\} \\ 
Researcher 1-F  & 1 & 0.256 & \{LMS Algorithm\} $\rightarrow$ \{Machine Learning\} \\
\hline \hline
\end{tabular}
} 
\end{center}
\end{table*}

\begin{table*}[tbp]
\caption{Total Matching Score on \cite{kayamori}}
\vspace{-3mm}
\label{tab:result-totalscore-kayamori}
\begin{center}
\scalebox{1}[1]{
\begin{tabular}{l|r|r|r}
\hline \hline
\multicolumn{1}{c|}{Researcher Name} & \multicolumn{1}{c|}{$\alpha = 0.5, \beta = 0.5$} & \multicolumn{1}{c|}{$\alpha = 0.8, \beta = 0.2$} & \multicolumn{1}{c}{$\alpha = 0.8, \beta = 0.2$} \\ \hline\hline
Researcher 1-A & 0.354 & {\bf 0.566} & 0.141  \\ 
Researcher 1-B & 0.304 & {\bf 0.486} & 0.121  \\ 
Researcher 1-C & {\bf 0.568} & {\bf 0.453} & {\bf 0.682} \\ 
Researcher 1-D & 0.175 & 0.280 & 0.070  \\ 
Researcher 1-E & 0.125 & 0.200 & 0.050 \\ 
Researcher 1-F & 0.278 & 0.111 & {\bf 0.444} \\
\hline \hline
\end{tabular}
} 
\end{center}
\end{table*}

\section{Synthesis Process of Evaluations}
\label{sec:SyntheisEvaluation}
As mentioned in the subsection \ref{sec:system}, the system has 2 kinds of evaluations. The recommendation candidates are determined according to each analysis procedure, there is the recommendation for only one researcher because the grant organization needs the letter of recommendation. That is, there is no two or more accepted applications in the university. Therefore, it is required for the staff to select one researcher from the candidates.

If the researcher contributes many kinds of research papers, the agreement degree to association rule becomes high. If the researcher selects good KAKEN keywords and proposes the KAKEN application, the matching score by TF-IDF becomes high. Of course, because such a researcher can acquire KAKEN grant, another grants are the state of being unnecessary for him. The achievement of acquisition of the grant is not easy and therefore the detailed adjustment of the recommendation is required by research administrations.

Based on such environments, we design the parameter to adjust two or more recommendation results. In this paper, the system outputs 2 kinds of recommendation lists. The parameter of the ratio of importance of the surface data and the deeply historical data is interactively defined while reviewing the calculation results. That is, if either agreement degree to the researcher is too far apart from other researchers, the recommendation list is applicable to select the researcher. However, in case of the degree are all alike, the system calculates to be weighted the agreement degree by using the defined parameter. In this paper, there are 2 kinds of agreement degrees of recommendation and we define the following equation.

\begin{equation}
(Total Score) =\alpha (Surface data) + \beta (Historical data),
\label{eq:total_score}
\end{equation}
\begin{equation}
\alpha + \beta = 1.0
\label{eq:total_score_paramter}
\end{equation}

\section{Experimental Results}
\label{sec:ExperimentalResults}
In this section, some matching results between about 240 researchers in Prefectural University of Hiroshima and 3 kinds of grants \cite{kayamori}-\cite{meiji} by our proposed system are described. Table \ref{tab:result-keyword-kayamori} shows the recommendation lists to the Kayamori Foundation \cite{kayamori} by using Surface data. The Kayamori Foundation is the grant about information science, and 5 researchers from A to E who belong to Department of Management and Information Systems were matched. Since the grant is related to Informatics research field, the researchers belonging to other departments were not matched. For the description of grant explanation (Surface data) as shown in Table \ref{tab:result-keyword-kayamori}, the sum of TF-IDF value for each KAKEN keyword by Eq.(\ref{eq:tfidf}) is normalized into the value $[0, 1]$. 
  
Table \ref{tab:result-rule-kayamori} shows the recommendation lists to the same grant by using historical data that mining the researcher's science papers, the summarized contents of the accepted research from the selected results, and the summarized application of the selected researcher from KAKEN database. The 2 researchers, C and F, were matched. Both the researcher C in Table \ref{tab:result-keyword-kayamori} and in Table \ref{tab:result-rule-kayamori} is the same person. Although the researcher F was not shown in \ref{tab:result-keyword-kayamori}, the researcher changes the main research field in KAKEN and the mining results from the surface data did not match the keywords in the Grant. \{X\} $\rightarrow$ \{Y\} in Table \ref{tab:result-rule-kayamori} is the acquired association rule. The sum of lift value for each matched association rule by using Eq.(\ref{eq:lift}) is normalized into $[0, 1]$ as `Historical data.'
  
By using the matching scores of the surface data and the historical data, the total matching score was calculated by Eq.(\ref{eq:total_score}). As shown in Table \ref{tab:result-totalscore-kayamori}, The 3 kinds of the parameter set of $\alpha$ and $\beta$ were prepared to investigate the appropriate ratio of $(\alpha, \beta)$: $(\alpha, \beta) = \{0.5, 0.5\}, \{0.8, 0.2\}, \{0.2, 0.8\}$. 

When we select the researchers to recommend, the order of researcher should be determined by using Table \ref{tab:result-totalscore-kayamori}. However, if there are many members in the list, the research administrator will be perplexed to select only one researcher. Therefore, the threshold value to total score in Eq.(\ref{eq:total_score}) was defined to make the good recommendation list. If the threshold value is $0.4$, the researcher 1-C was only recommended in case of $\alpha = 0.5, \beta = 0.5$. In case of $\alpha = 0.8, \beta = 0.2$ and $\alpha = 0.2, \beta = 0.8$, the recommended researcher is 1-A, 1-B, 1-C and 1-C, 1-F, respectively. The overall results shows the researcher 1-C keeps high score for all parameter settings. In this way, the appropriate parameter is defined as the 2 or less researchers are selected in the list. Table \ref{tab:result-totalscore-yazuya} and Table \ref{tab:result-totalscore-meiji} show the the total matching score for the other grant information \cite{yazuya} and \cite{meiji}, respectively.

Table \ref{tab:result-matching} shows the overview of matching results for 3 grants \cite{kayamori}-\cite{meiji}. The column `No. of Surface data' and `No. of Historical data' are the number of researchers recommended by KAKEN keywords and Association rules, respectively. The column `No. of Total Score' is the number of strong recommended researchers when $\alpha = 0.5, \beta = 0.5$. For all the matching results, only one or two researcher were strongly recommended by the system. 

\begin{table*}[tbp]
\caption{Total Matching Score on \cite{yazuya}}
\vspace{-3mm}
\label{tab:result-totalscore-yazuya}
\begin{center}
\scalebox{1}[1]{
\begin{tabular}{l|r|r|r}
\hline \hline
\multicolumn{1}{c|}{Researcher Name} & \multicolumn{1}{c|}{$\alpha = 0.5, \beta = 0.5$} & \multicolumn{1}{c|}{$\alpha = 0.8, \beta = 0.2$} & \multicolumn{1}{c}{$\alpha = 0.8, \beta = 0.2$} \\ \hline\hline
Researcher 2-A & 0.219 & 0.350 & 0.088\\
Researcher 2-B & 0.078 & 0.124 & 0.031\\
Researcher 2-C & 0.143 & 0.229 & 0.057\\
Researcher 2-D & {\bf 0.633} & {\bf 0.729} & {\bf 0.536}\\
Researcher 2-E & 0.086 & 0.137 & 0.034\\
Researcher 2-F & 0.246 & 0.393 & 0.098\\
Researcher 2-G & 0.141 & 0.225 & 0.056\\
Researcher 2-H & {\bf 0.430} & 0.393 & {\bf 0.468}\\
Researcher 2-I & 0.233 & 0.372 & 0.093\\
Researcher 2-J & 0.154 & 0.246 & 0.062\\
Researcher 2-K & 0.175 & 0.280 & 0.070\\
Researcher 2-L & 0.175 & 0.280 & 0.070\\
Researcher 2-M & 0.180 & 0.288 & 0.072\\
Researcher 2-N & 0.394 & 0.158 & {\bf 0.630}\\
Researcher 2-O & 0.225 & 0.090 & 0.360\\
Researcher 2-P & 0.161 & 0.064 & 0.258\\
Researcher 2-Q & 0.161 & 0.064 & 0.258\\
Researcher 2-R & 0.161 & 0.064 & 0.258\\
Researcher 2-S & 0.157 & 0.063 & 0.251\\
Researcher 2-T & 0.157 & 0.063 & 0.251\\
Researcher 2-U & 0.147 & 0.059 & 0.234\\
Researcher 2-V & 0.122 & 0.049 & 0.194\\
\hline \hline
\end{tabular}
} 
\end{center}
\end{table*}

\begin{table*}[tbp]
\caption{Total Matching Score on \cite{meiji}}
\vspace{-3mm}
\label{tab:result-totalscore-meiji}
\begin{center}
\scalebox{1}[1]{
\begin{tabular}{l|r|r|r}
\hline \hline
\multicolumn{1}{c|}{Researcher Name} & \multicolumn{1}{c|}{$\alpha = 0.5, \beta = 0.5$} & \multicolumn{1}{c|}{$\alpha = 0.8, \beta = 0.2$} & \multicolumn{1}{c}{$\alpha = 0.8, \beta = 0.2$} \\ \hline\hline
Researcher 3-A & 0.208 & 0.332 & 0.083\\
Researcher 3-B & {\bf 0.586} & {\bf 0.483} & {\bf 0.688}\\
Researcher 3-C & 0.303 & {\bf 0.485} & 0.121\\
Researcher 3-D & 0.302 & {\bf 0.482} & 0.121\\
Researcher 3-E & 0.207 & 0.331 & 0.083\\
Researcher 3-F & 0.225 & 0.090 & 0.360\\
\hline \hline
\end{tabular}
} 
\end{center}
\end{table*}

\begin{table*}[tbp]
\caption{Overview of Matching Results}
\vspace{-3mm}
\label{tab:result-matching}
\begin{center}
\scalebox{1}[1]{
\begin{tabular}{l|r|r|r}
\hline \hline
Foundation & No. of Surface data  & No. of Historic data & No. of Total Score\\ \hline\hline
Kayamori Foundation \cite{kayamori}        & 5  & 2 & 1\\
Yazuya Co., Ltd. \cite{yazuya}                   & 13 & 11 & 2\\
Meiji Yasuda Life Foundation \cite{meiji}  &  5 & 3 & 1\\
\hline \hline
\end{tabular}
} 
\end{center}
\end{table*}

\section{Conclusion}
\label{sec:Conclusion}
This paper explains our developed Grants-in-Aid system and some modification about the retrieval algorithms. The system can recommend the researchers related to each grant information in terms of 2 kinds of analysis parts, that is the surface data based on KAKEN keywords, and the historical data based on association rules. In the experimental results, one or two researchers were strongly recommended to each grant organization. However, the parameter setting $\alpha$ and $\beta$ is very important and the expected matching results may not be acquired according to these value. In future, we will improve the performance of the developed system via operation and develop the method to adjust the parameter $\alpha$ and $\beta$ automatically according to the acquired recommendation lists.

\end{document}